\documentclass[preprint2]{aastex62}
\usepackage[utf8]{inputenc}
\usepackage{graphicx}
\usepackage{wrapfig}
\usepackage{array}
\usepackage{tabularx}
\usepackage{gensymb}
\usepackage{lineno}

\begin{document}

\title{Evidence that a novel type of satellite wake might exist in Saturn's E ring}

\author{M.M. Hedman}
\affil{Department of Physics, University of Idaho, Moscow Idaho 83843}
\author{M. Young}
\affil{Department of Physics, University of Idaho, Moscow Idaho 83843}

%\linenumbers
\begin{abstract}
Saturn's E ring consists of micron-sized particles launched from Enceladus by that moon's geological activity. A variety of small-scale structures in the E-ring's brightness have been attributed to tendrils of material recently launched from Enceladus. However, one of these features occurs at a location where Enceladus' gravitational perturbations should concentrate background E-ring particles into structures known as satellite wakes. While satellite wakes have been observed previously in ring material drifting past other moons, these E-ring structures would be the first examples of wakes involving particles following horseshoe orbits near Enceladus' orbit. The predicted intensity of these wake signatures are particularly sensitive to the fraction E-ring particles' on orbits with low eccentricities and semi-major axes just outside of Enceladus' orbit, and so detailed analyses of these and other small-scale E-ring features should place strong constraints on the orbital properties and evolution of E-ring particles.

\vspace{.5in}

\end{abstract}

\section{Introduction}

Saturn's E ring is a broad and tenuous ring composed of very small ($<5\mu$m) ice grains that encompasses the orbits of many of Saturn's icy moons \citep[see][for recent reviews] {Kempf18, Hedman18}. This ring is brightest around the orbit of the small moon Enceladus, and so it was long suspected that it had some connection with that moon \citep{Show91, Horanyi92, HB94, BHS01}. This suspicion was dramatically confirmed by the Cassini mission, which revealed that Enceladus was geologically active, with a series of fissures around the moon's south pole launching micron-sized particles from beneath the moon's surface into orbit around Saturn \citep[see][and references therein]{Schenk18}. 

In addition to revealing the E-ring's origin, the Cassini mission clarfied several aspects of the ring's structure and dynamics. Theoretical models of the E-ring's large-scale structure suggested that its broad extent arises because the orbital eccentricities and inclinations of the small particles launched from Enceladus can be greatly increased by a combination of solar radiation pressure and Saturn's offset magnetic field, while on longer timescales the particles are slowly eroded and transported outwards by plasma drag \citep{Horanyi92, Hamilton93, Hamilton93b, HB94, Juhasz02, Juhasz04, Juhasz07, Horanyi08}. Cassini observations have confirmed many aspects of these models, but have also identified large-scale asymmetries in the ring's brightness which imply that particle transport within this ring is more complex than expected \citep{Hedman12, Ye16}. 

\begin{figure*}[p]
    \centering
    \resizebox{6.5in}{!}{\includegraphics{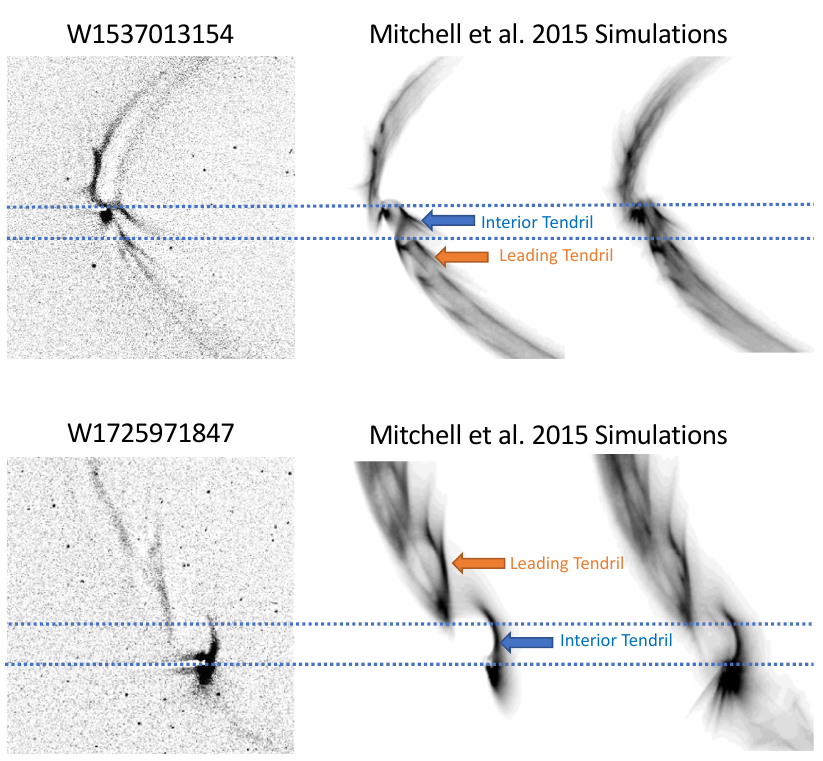}}
    \caption{Comparisons of E-ring images with numerical simulations, adapted from \citet{Mitchell15}. The left panels show high-pass filtered and brightness inverted versions of two images of the E ring around Enceladus. The middle and right panels show two different simulations of material recently ejected from Enceladus, with different assumptions about Enceladus' activity. Two concentrations of material near Enceladus are marked as the Interior and Leading Tendrils. (Note the direction of orbital motion is counterclockwise in the top image and clockwise in the bottom image). For both observations, the leading tendril extends closer to Enceladus in the images than it does in the simulations, so that it is more prominent between the two dotted horizontal lines.}
    \label{m15}
\end{figure*}

Fine-scale structures in the E ring close to Enceladus provide important opportunities to further explore the connections between Enceladus and the E ring. Thus far, published studies of these features have focused on whether they can be attributed to the activity of specific sources near Enceladus' south pole \citep{Kempf10, Mitchell15}. 
However, in this paper we will take a closer look at a sub-set of these features that might represent structures in the background E ring rather than freshly ejected material from Enceladus. 

Figure~\ref{m15} shows high-pass filtered and brightness-inverted versions of two images of the E ring first analyzed by \citet{Mitchell15}, along with numerical simulations performed by those authors that track particles of various sizes launched from various locations around Enceladus' south pole. The particles in the simulations are organized into elongated structures called ``tendrils''. One of these tendrils extends interior from the moon and is designated as the interior tendril, while others are found at various locations in front of the moon, with the first of the ``leading tendrils'' being the most prominent in the images. If we take a closer look at the first leading tendril, we can note that it extends closer to Enceladus in the images than it does in the simulations. This is most obvious in the earlier image W1537013154, where a prominent band can be seen between the two dashed lines that is not present in the simulations. For W1725971847 the difference is more subtle, but it still appears that the intensity of the band remains constant across the dotted line, whereas the simulations predict a sharp edge to this feature. In principle, these discrepancies could potentially be resolved by simulating plume particles with a broader range of physical parameters. However, it turns out that these features could also be density variations in pre-existing E-ring particles known as satellite wakes.

Satellite wakes have been observed in Saturn's main rings near the small moons Pan and Daphnis \citep{Showalter86, Horn96, Tiscareno07}. These wakes arise because the gravitational perturbations from a moon produce organized epicyclic motions in nearby ring particles, which in turn produce localized variations in the surface density with a similar orientation to the features seen here \citep{Murray01}. 

\citet{Mitchell15} argued that it was unlikely for any E-ring features to represent wakes because the spacing of the features was inconsistent with those expected for standard wakes, and because it is unlikely such structures would form within a ring where most of the particles have substantial orbital eccentricity. However, a tenuous ring like the E ring can contain a novel class of satellite wakes that occur in material following horseshoe orbits. These ``rebound wakes'' occur at very different locations from other satellite wakes, and the most prominent of these wakes should occur very close to the extensions of the leading tendril seen in these two images. 

In the rest of this paper, we argue that satellite wakes are a reasonable potential explanation for these specific features, and so this possibility merits further examination with additional numerical simulations. In Section~\ref{theory} we use Hill's Equations to demonstrate that material on horseshoe orbits can produce wake-like density variations. In Section~\ref{methods} we describe how we process the two images in order to determine the locations and brightnesses of the relevant features. In Section~\ref{results} we show that the bright features extending beyond the tips of the leading tendrils are consistent with the expected locations of these wakes. We also show that there could be enough particles on the appropriate orbits to produce the observed brightness variations. Finally, in Section~\ref{discussion} we discuss the potential implications of this feature being composed of particles on horseshoe orbits for the orbital distribution of E-ring particles. 

\begin{figure*}
\resizebox{6.5in}{!}{\includegraphics{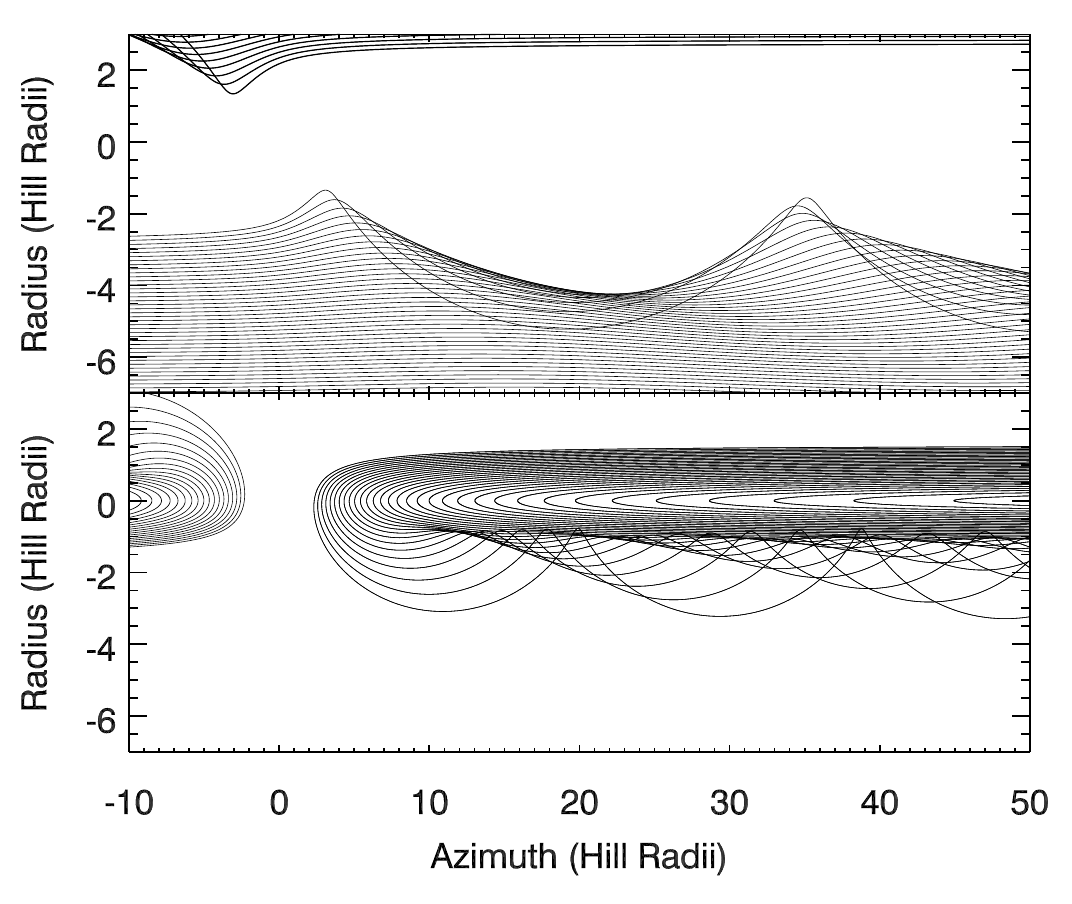}}
\caption{Trajectories of particles on orbits with the same eccentricity and pericenter location as  a satellite (located at the origin), computed using Hill's Equations. The top panel shows the trajectories of particles on that pass by the moon, while the bottom panel shows the trajectories of particles that undergo horseshoe motion. Note that the initial semi-major axis spacing of the trajectories is different in the two panels in order to better visualize the locations of the relevant wake patterns.}
\label{models}
\end{figure*}

. 

\section{Theory of satellite wakes in material on horseshoe orbits}
\label{theory}

The basic theory of satellite wakes was first developed to explain periodic brightness and optical depth variations found in the vicinity of the narrow Encke Gap in Saturn's main rings \citep{Showalter86}, and the relevant physics can be illustrated using Hill's Equations.

Consider a reference frame whose origin is at the location of a satellite of mass $m_s$ on an orbit with semi-major axis $a_s$ around a planet of mass $M_P$. Let the coordinates $\xi$ and $\eta$ be aligned with the local radial and azimuthal directions, measured in units of the satellite's Hill Radius:
\begin{equation}
R_H = a_s(m_s/3M_P)^{1/3}.
\end{equation}
Note that for Enceladus the Hill Radius is $ 939.74$ km. In this coordinate system, the equations of motion for a test particle are given by Hill's Equations \citep{HenonPetit86}:
\pagebreak
\begin{equation}
    \ddot{\xi}-2n\dot{\eta}=3n^2\xi\left(1-\frac{1}{(\xi^2+\eta^2)^{3/2}}\right)
\end{equation}
\begin{equation}
    \ddot{\eta}+2n\dot{\xi}=-\frac{3n^2\eta}{(\xi^2+\eta^2)^{3/2}}
\end{equation}
where $n$ is the satellite's orbital mean motion. Note these equations describe the motion of the test particle relative to the satellite even if the satellite is on an eccentric orbit \citep{HenonPetit86}.

Standard satellite wakes arise when we consider particles initially on orbits with fixed $\xi$ interior or exterior to the moon and when the semi-major axis difference between the particles and the moon $\Delta a$ is sufficiently large that the particles will not undergo horseshoe motion (see top panel of Figure~\ref{models}), Due to Keplerian shear, these particles all drift in longitude relative to the satellite. As the particles drift past the moon, the gravitational pull from the moon causes all the particles to acquire finite eccentricities and thus execute radial motions with an amplitude that steadily decreases with increasing $\Delta a$ \citep{JT66, LP79, Murray01}. As these particles continue to move downstream from the moon, their radial epicyclic motion causes them to follow quasi-sinusoidal paths with wavelength $3\pi\Delta a$ \citep{Murray01}. These systematic trends in the wavelengths of the particle's trajectories in turn cause the distances between particles on nearby semi-major axes to vary, yielding variations in the local particle density. More specifically, there are diagonal bands where the particle trajectories get very close together and so the density should be exceptionally high. The first of these wakes is most obvious between 5 and 25 Hill Radii in azimuth and -2 to -4 Hill Radii in radius, while the second wake starts at 35 Hill Radii and extends off the page. 

Material orbiting on semi-major axes closer to the satellite is not normally considered suitable for forming satellite wakes because instead of passing the moon, the particles execute horseshoe motion due to the semi-major axis evolution induced by the satellite's gravitational perturbations. However, as shown in the bottom panel of Figure~\ref{models}, wake-like structures can occur in this material. In this case, particles approach the moon on initially circular orbits from the right, but turn around before reaching the satellite's location at distances that depend on the initial semi-major axis separation. When the initial semi-major axis difference between the particle and the moon is large enough the particle also acquires a significant eccentricity as it turns around in front of the moon. The resulting radial motions again cause the particles to follow oscillating trajectories with wavelengths that are proportional to the semi-major axis separation from the moon, giving rise to diagonal bands of enhanced density similar to those seen for material passing by the moon. To distinguish these wakes from the classic satellite wakes, we call the wakes in material that drifts past the moon ``passing wakes'' and the wakes in the material undergoing horseshoe motion ``rebound wakes''. 

While the locations of both passing and rebound wakes can be most easily illustrated with the trajectories of particles initially on orbits with a constant $\xi$, most of the particles in the E ring are not on these sorts of orbits. Trajectories with constant $\xi$ in the E ring correspond to particles with the same orbital eccentricity and pericenter as Enceladus \citep{HenonPetit86}, and the particles in the E ring have a wide range of eccentricities that are driven by  non-gravitational forces like solar radiation pressure  \citep{Horanyi92, Hamilton93, HB94, Juhasz02, Juhasz04, Juhasz07, Horanyi08, Hedman12, Ye16}.  Particles with orbital eccentricities and pericenters that differ from those of Enceladus would  follow cycloid-like trajectories spanning a range of $\xi$ values. For such particles, it is  difficult to visualize the variations in the particle density using individual particle trajectories. Instead one needs to generate maps of particle density for each specified distribution of orbital elements by simulating a sufficiently large number of particles. \citet{Mitchell15} did two such simulations, one where all the particles had negligible eccentricity and another where the particles had eccentricities of 0.02. The wake signals were not obvious in the latter simulation, implying that only particles on nearly circular orbits can produce wake signatures. However, the features seen in the images are actually rather subtle variations in the E-ring's brightness, being only about 5\% brighter than their surroundings (see Section 4.2). Hence it is worth examining how the locations and intensity of potential wake signatures vary among particle populations with finite eccentricities.

The spatial density of particles with a range of orbital properties can be derived from the trajectories of individual particles computed using Hill's Equations. However, for  any given initial particle orbital parameter distribution only a finite number of these tracks can be calculated, and this will give rise to artificial small-scale density variations that can obscure the desired wake signatures. This ``sampling noise'' can always be reduced by increasing the number of simulated trajectories, but this will also increase the time needed to do the simulations. {In practice, we find that 10$^5$-10$^6$ trajectories need to be simulated in order for wake-like signatures with amplitudes of 5\% to be detectable.} These sorts of calculations are therefore an inefficient way to explore a broad parameter space of potential orbital parameter distributions. Hence we will instead consider a small number of situations that illustrate important aspects of the signals from satellite wakes.

First, we will simulate particles with a limited range of semi-major axes and eccentricities to demonstrate that the locations of the rebound wake signatures are insensitive to the orbital parameter distribution, and their intensity is proportional to the fraction of particles on low-eccentricity orbits with the appropriate range of semi-major axes. For these simulations we  trace the trajectories of 10$^5$ particles randomly  drawn from a population with the following distributions of initial orbital parameters:
\begin{itemize}
\item A uniform distribution of initial eccentricities between 0 and 0.04 (corresponding to initial radial oscillations with amplitudes between 0 and 10 Hill Radii).
\item A uniform distribution of pericenter locations. 
\item A Gaussian distribution of semi-major axes with a center 1.5 Hill Radii external to the moon's orbit and a Gaussian width of 0.5, 1.0 or 2.0 Hill radii. 
\end{itemize}
Note that the eccentricity and semi-major axis distributions considered here are {\bf not} intended to be fully representative of the E-ring, but instead serve to illustrate how the locations and intensity of wake features depend on the orbital properties of the ring particles. The eccentricity distribution was chosen because the wake signatures see in Figure~\ref{models} are only likely to be prominent for particles on orbits with initial radial motions  less than 1 Hill Radii, which corresponds to eccentricities less than about 0.005. The above eccentricity distribution therefore ensures that the fraction of particles on such orbits is small but not zero. Furthermore, this distribution gives a median eccentricity similar to that used by \citet{Mitchell15}. The pericenter distribution is uniform  both for the sake of simplicity and because the observed large-scale structure of the E ring does not show strong asymmetries in pericenter locations \citep{Hedman12, Ye16}.  Finally, the semi-major axis distributions are chosen so that there are some particles on orbits likely to produce rebound wakes in all three cases, but the fraction of particles on these orbits differs.

\begin{figure}
    \centering
    \resizebox{3.4in}{!}{\includegraphics{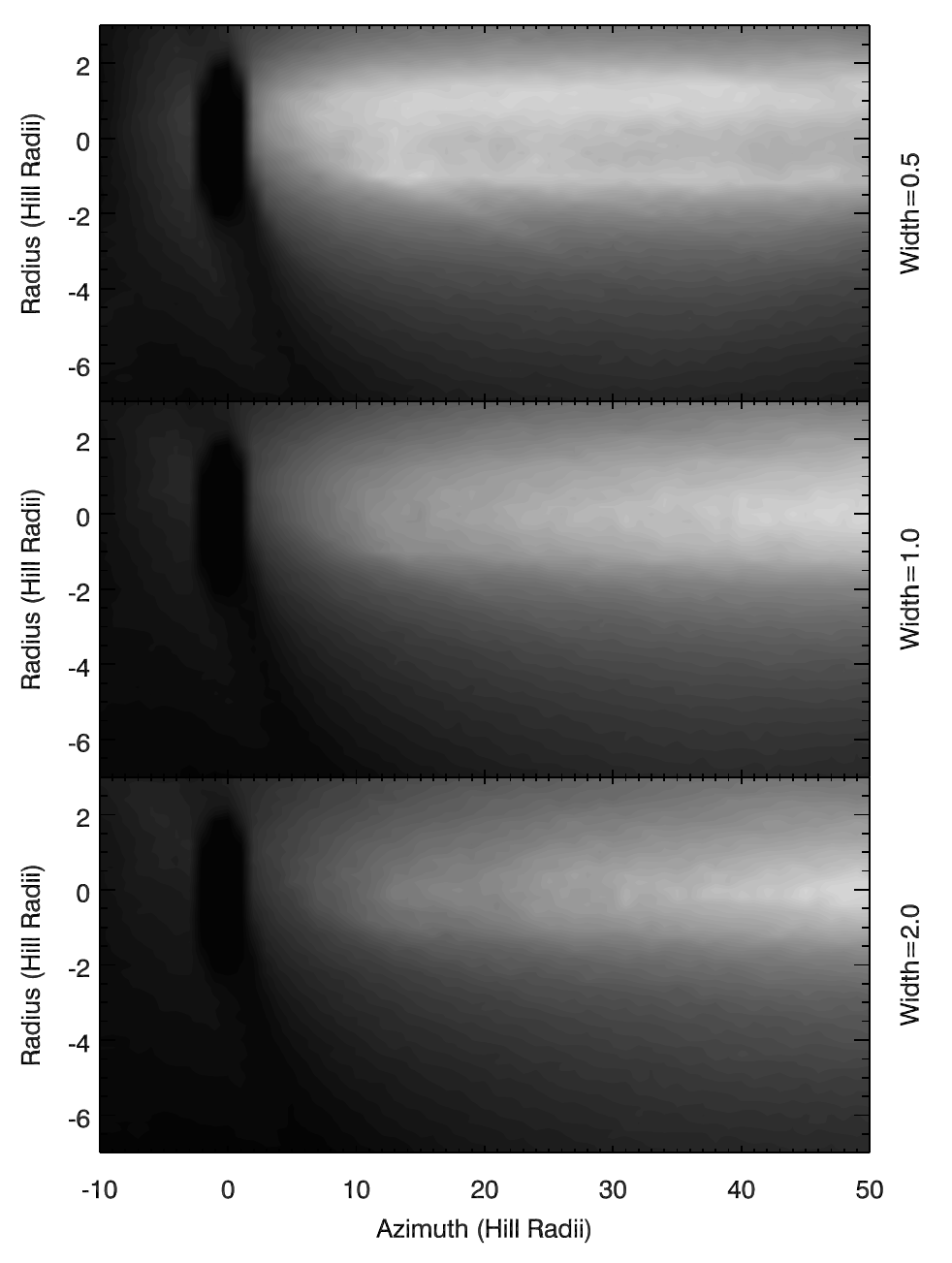}}
    \caption{Simulated particle density maps for particle populations with uniform eccentricity distributions between 0 and 0.04, uniform pericenter location distributions and Gaussian semi-major axis distributions with a center at +1.5 Hill Radii and the three different widths indicated. All panels show a broad distribution of ring particles, but in the top panel a density enhancement can be seen at the expected location of the rebound wake around -1 Hill Radii in radius and 15 Hill Radii in longitude. Note that no passing wakes are visible in this simulation because of the assumed semi-major axis distribution.}
    \label{hillsim}
\end{figure}

\begin{figure}
    \centering
    \resizebox{3.5in}{!}{\includegraphics{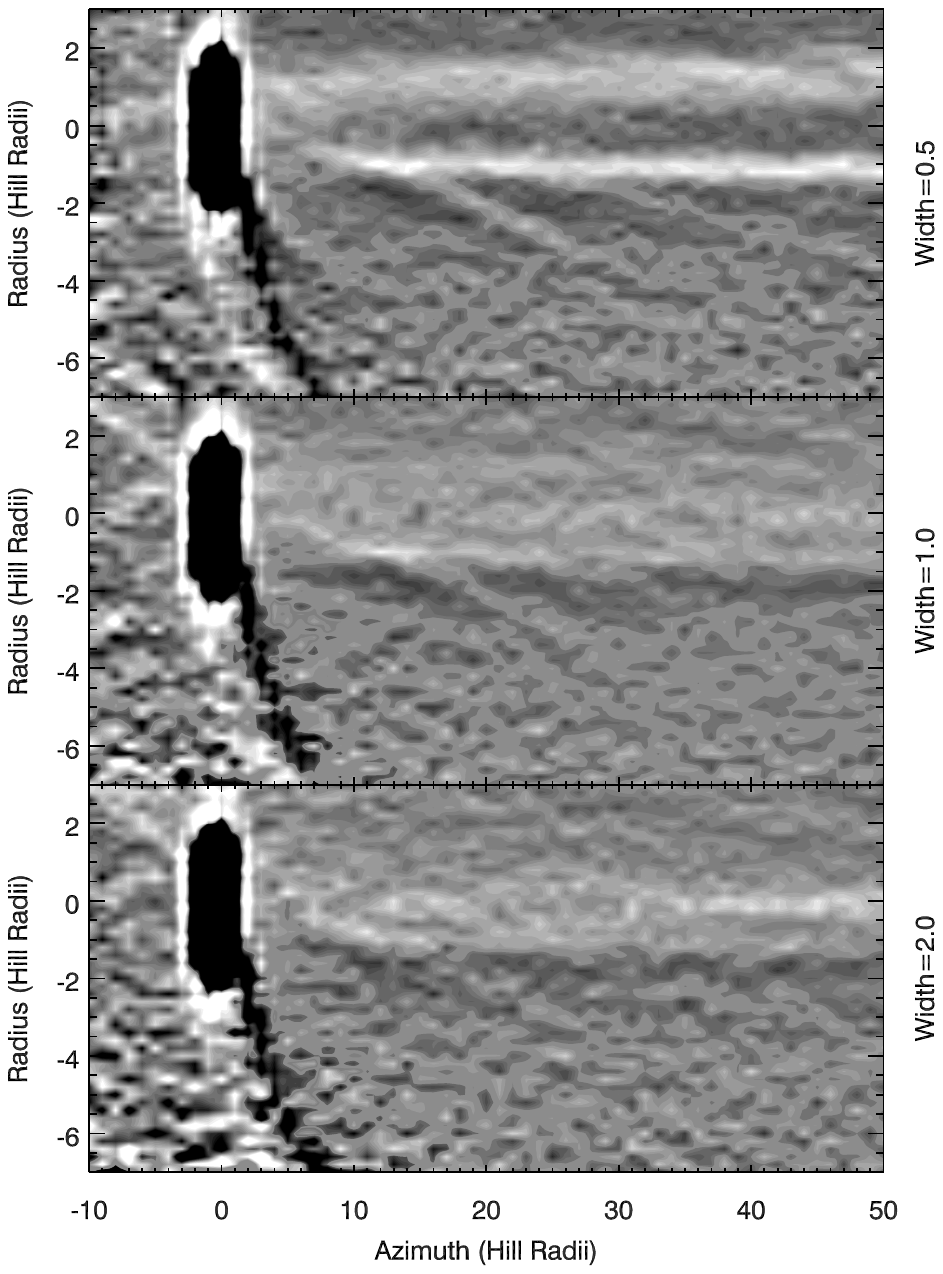}}
    \caption{The same simulated particle density maps as shown in Figure~\ref{hillsim}, except they have been high-pass filtered by removing a smoothed version of the maps with a smoothing length of 1.5 Hill Radii.  The brightness range corresponds to fractional brightness variations of $\pm$10\%. Here we can see a brightness maximum due to the  rebound wake in all three simulations. Note the strength of the feature decreases as the width of the semi-major axis distribution increases.}
    \label{hillsim2}
\end{figure}

The trajectories for the 10$^5$ particles drawn from each of these populations were evaluated at fixed time steps using Hill's equations, starting from a fixed longitude over 70 Hill Radii in front of the moon. These trajectories were terminated if the particles got within 1 Hill Radius of the moon. The resulting tables of particle locations were then binned together to estimate the relative density of particles as functions of radius and longitude relative to the moon. The resulting density maps are shown in Figure~\ref{hillsim}. These maps are clearly dominated by radial density variations that mostly reflect the initial eccentricity distribution of the particles. The simulations with larger spreads in initial semi-major axis also show a gradual decrease in the peak particle density near the moon's location. This is because a substantial fraction of the particles are on horseshoe orbits that turn around at various distances from the moon. For the simulation with the tightest semi-major axis distribution, there is a $\subset$ shape in the locations of peak particle density that corresponds to the expected track of circular particles with initial semi-major axes around 1.5 Hill Radii outside of the moon's orbit. The lower branch of this track is narrower than the upper branch because some fraction of the particles either pass by the moon or are lost when they get too close to the moon.

Turning our attention to potential  signals from rebound wakes, we can note that the top panel in Figure~\ref{hillsim} shows a bright diagonal streak running between  -1 and -3 Hill Radii in radius and 10 and 20 Hill Radii in longitude. This streak falls at the expected location of the first rebound wake shown in Figure~\ref{models},  so signatures of rebound wakes can indeed appear in particle populations with a range of eccentricities and semi-major axes. Similar wake signatures  are also present in the other two simulations, but they are harder to see against the background trends. Hence Figure~\ref{hillsim2} shows the fractional difference between each of these maps and versions of the maps smoothed over a scale of 1.5 Hill Radii. In all of these high-pass-filtered maps, there is a localized density enhancement around -1 Hill Radii in radius and 12 Hill Radii in longitude. These density enhancements all occur where the first rebound wake is expected to be the most intense, so it is reasonable to attribute them all  to this wake. The location of this density enhancement therefore appears to be insensitive to the particles' orbit distribution. 
 
The intensity of these wake signatures also clearly declines as the width of the semi-major axis distribution increases. The fractional density variations associated with this feature are 10\%,  5\% and 2.5\% for these three simulations. These numbers are comparable to the fraction of particles in the simulations on initial orbits with eccentricities below 0.005 and semi-major axes between 1 and 1.5 Hill radii exterior to the moon. Thus it is reasonable to conclude that particles on these particular orbits generate a wake signature at the location predicted by Figure~\ref{models}, while particles on orbits with higher eccentricities and different semi-major axes simply contribute a smooth background in this region. This naturally results in the observed fractional density (or brightness) variations due to the wake being roughly equal to the fraction of particles on suitable orbits in this region because the density variations due to the wake are of order unity for the particles with the appropriate orbital parameters.

\begin{figure}
    \centering
    \resizebox{3.4in}{!}{\includegraphics{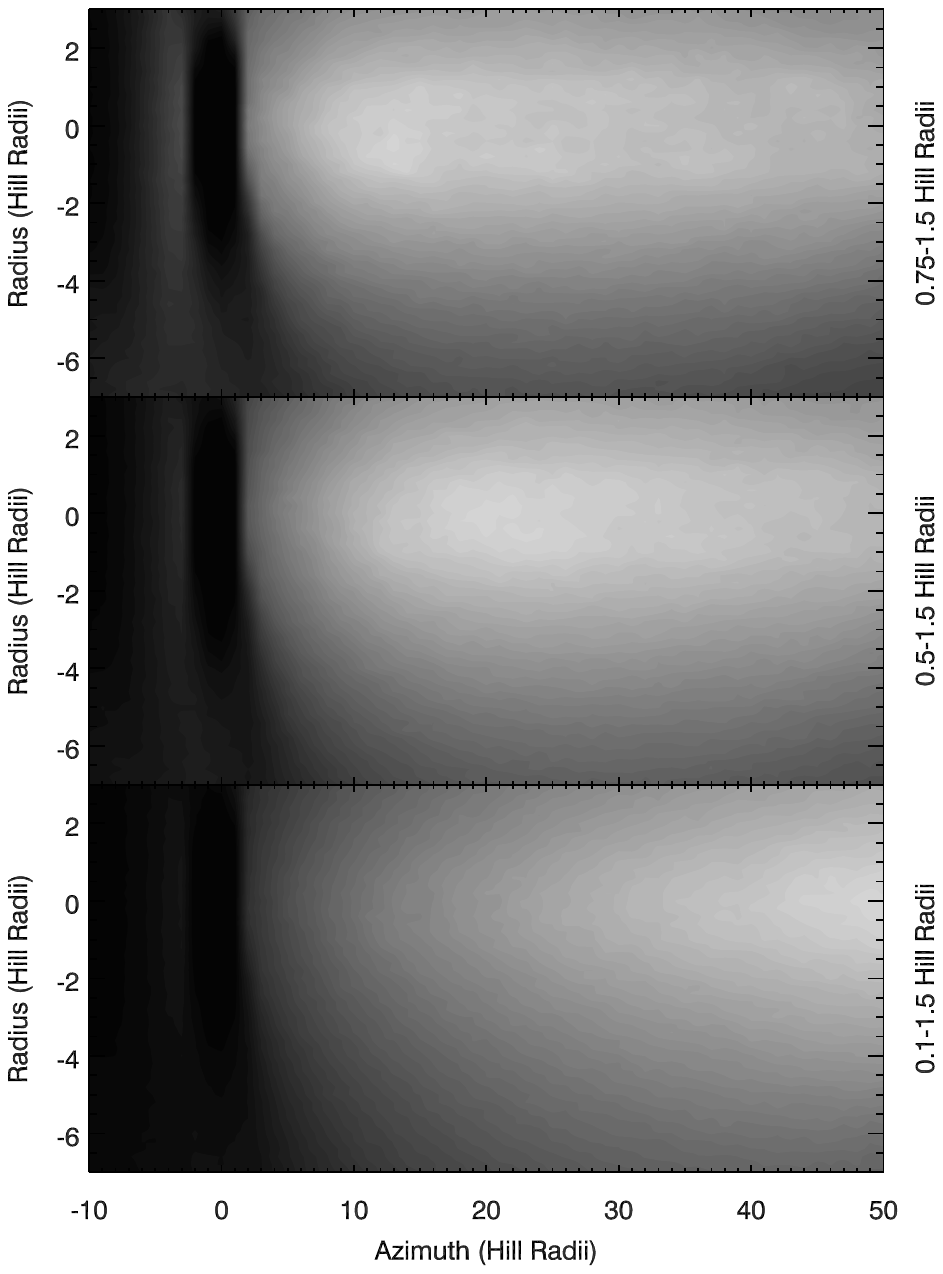}}
    \caption{{Simulated particle density maps for particle populations with the \citet{Hedman20} eccentricity distributions,  uniform pericenter location distributions and uniform semi-major axis between the values indicated. All panels show a broad distribution of ring particles. Note that the simulated brightness distributions for all three populations are very similar, although a weak rebound wake signal can be seen in the top panel.}}
    \label{hillsim3}
\end{figure}

\begin{figure}
    \centering
    \resizebox{3.4in}{!}{\includegraphics{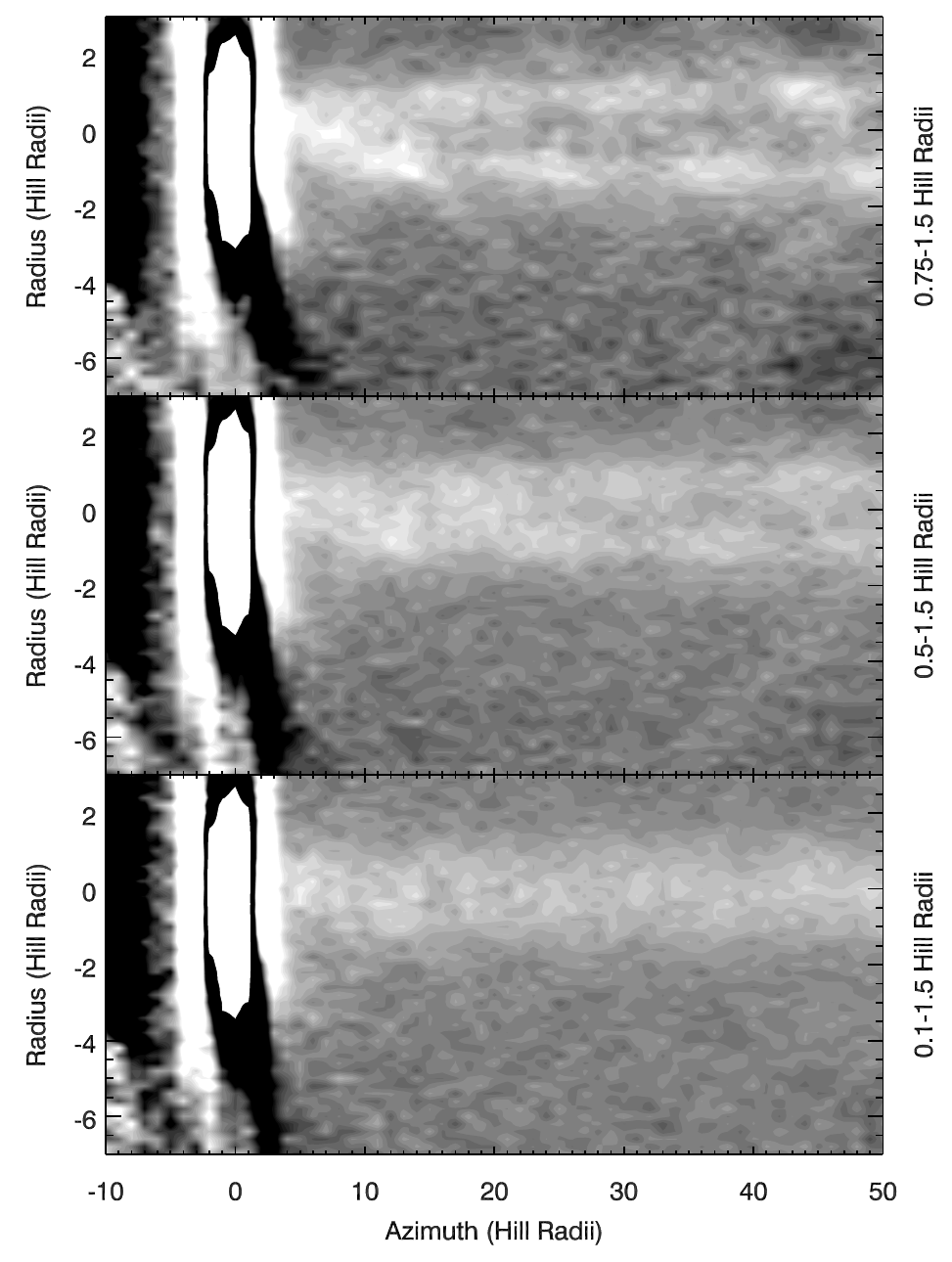}}
    \caption{{The same simulated particle density maps as shown in Figure~\ref{hillsim3}, except they have been high-pass filtered by removing a smoothed version of the maps with a smoothing length of 4 Hill Radii.  The brightness range corresponds to fractional brightness variations of $\pm$10\%. Here we can see a brightness maximum due to the rebound wake around +12 Hill Radii in azimuth and -1 Hill radii in Radius in at least the top two simulations. Again, the brightness of this feature decreases as the width of the semi-major axis distribution increases.}}
    \label{hillsim4}
\end{figure}

{While the above simulations suggest that the amplitude of the wake signatures have a relatively straightforward relationship with the particles' orbit element distribution, complications arise when we consider particle populations with broad eccentricity distributions like the E ring.  \cite{Hedman20} found that the radial brightness profile of the E ring could be reasonably well matched if the particles in the E ring had the following eccentricity distribution.
\begin{equation}
\mathcal{F}(e)=\frac{\mathcal{F}_0}{1+e^2/e_o^2}[1-\exp(-e/e_c)]  
\label{fe}  
\end{equation}
with $e_o\simeq 0.17$ and $e_c \simeq 0.01$. Such an eccentricity distribution has only about 0.5\% of the particles on orbits with eccentricities less than 0.005. At first, this would appear to imply that there would not be sufficient particles to produce wake signals of intensities comparable to those seen in the images and the simulations shown in Figure~\ref{hillsim2}.  However, particles with small eccentricities spend more time close to Enceladus' orbit than particles with large eccentricities, so the particle density near Enceladus' orbit will have a larger fraction of particles on low-eccentricity orbits. The density of particles near the E-ring's core with a specified range of eccentricities can be estimated using the following formula \citep{Hedman20}:
\begin{equation}
d(r)=\int_{e_{min}}^{e_{max}}\frac{r\mathcal{F}(e)}{\pi a^2\sqrt{e^2-(r/a-1)^2}}de.    
\end{equation}
This formula indicates that  around Enceladus' orbit roughly 10\% of the particles have orbital eccentricities less than 0.005, and a similar fraction should be within 0.005 of Enceladus' orbital parameters. This would imply that even a ring with a broad eccentricity distribution could produce  wake signals comparable to those observed in the images (provided a sufficiently high fraction of the particles on low-eccentricity orbits had the appropriate semi-major axes).}

{In order to demonstrate that wake signals can still be reasonably prominent in particle populations with broad eccentricity distributions, we  performed simulations tracing the trajectories of 10$^6$ particles randomly  drawn from the a population with the following distributions of initial orbital parameters:
\begin{itemize}
\item The eccentricity distribution given by Equation~\ref{fe}.
\item A uniform distribution of pericenter locations. 
\item A uniform distribution of semi-major axes between +0.1, 0.5 or 0.75 Hill Radii and 1.5 Hill Radii of Enceladus
\end{itemize}
As with the previous simulations, these trajectories were evaluated at fixed time steps using Hill's equations, starting from a fixed longitude over 70 Hill Radii in front of the moon, and these trajectories were terminated if the particles got within 1 Hill Radius of the moon.  Again, it is important to note that these calculations are not meant to be a fully realistic simulation of the E ring. In particular, Hill's equations are probably not accurate equations of motion for the particles with high eccentricities at large distances from the moon's orbit. Forutnately, since we are only concerned with the particle distribution near Enceladus' orbit here, and particles on large eccentricities only contribute to the relatively uniform background density in this region, such inaccuracies should not significantly affect the intensity of the wake signatures in these simulations. Also, as before, we deliberately chose semi-major axis distributions that would contain sufficient particles on suitable orbits to produce wake signatures.}

{Figure~\ref{hillsim3} shows the resulting density maps for these simulations. As expected, the overall brightness trends in these simulations are smoother than the ones shown in Figure~\ref{hillsim}, and the differences between the three simulations are more subtle. More importantly, a wake signature is faintly visible in the  simulation with the narrowest semi-major axis distribution. This signal can be seen much more clearly in Figure~\ref{hillsim4}, which shows the same simulations after high-pass filtered by subtracting a version of the map smoothed over 4 Hill Radii. In these filtered maps, the simulation with the narrowest semi-major axis distribution shows a clear brightness enhancement around -1 Hill Radius in radius and 12 Hill Radii in longitude, consistent with the expected location of a rebound wake. A weak signal around the same location can also be seen in the middle panel. For the bottom simulation the signal is too weak to be robustly detectable, but there may be a slight brightness enhancement near the noise level at the appropriate location. The peak amplitudes of the wake signals in these simulations are around 6\%, 3\% and  1.5\%, respectively, which is again comparable to the brightness variations associated with the small-scale structures in the E-ring. These simulations therefore demonstrate that detectable wake signals can occur in particle populations with broad eccentricity distributions.}

\section{Observed properties of the bright streaks in the E ring}
\label{methods}

\begin{table*}
\caption{Parameters for images used in this study.}
\hspace{-1in}
{\resizebox{8in}{!}{
\begin{tabular}{|c|c|c|c|c|c|c|c|}
\hline \hline
 Image ID & Filter & Date & Phase Angle  & Sub-solar 
 & Enceladus & Enceladus Orbital & Pixel Scale  \\
 & & & (degrees) & Longitude (deg.) & Longitude (deg.) &
 Pericenter Long. (deg.) & (km/pixel) \\
\hline
 W1537013154 & Clear & 2006-258T11:34:10 & 174.4 & -169.8 & -72.1 & -81.6 & 141.9 \\
 W1752971847 & Red & 2013-200T23:41:28 & 169.7 & -91.7 & -12.3 & 44.9 & 72.5 \\ 
\hline \hline 
\end{tabular}}}
\label{images used}
\end{table*}

\begin{figure*}
\resizebox{7in}{!}{\includegraphics{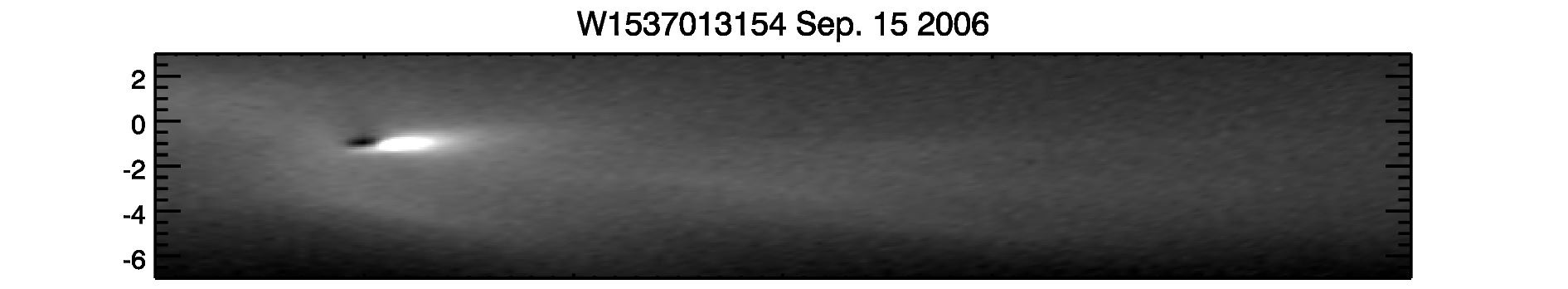}}
\resizebox{7in}{!}{\includegraphics{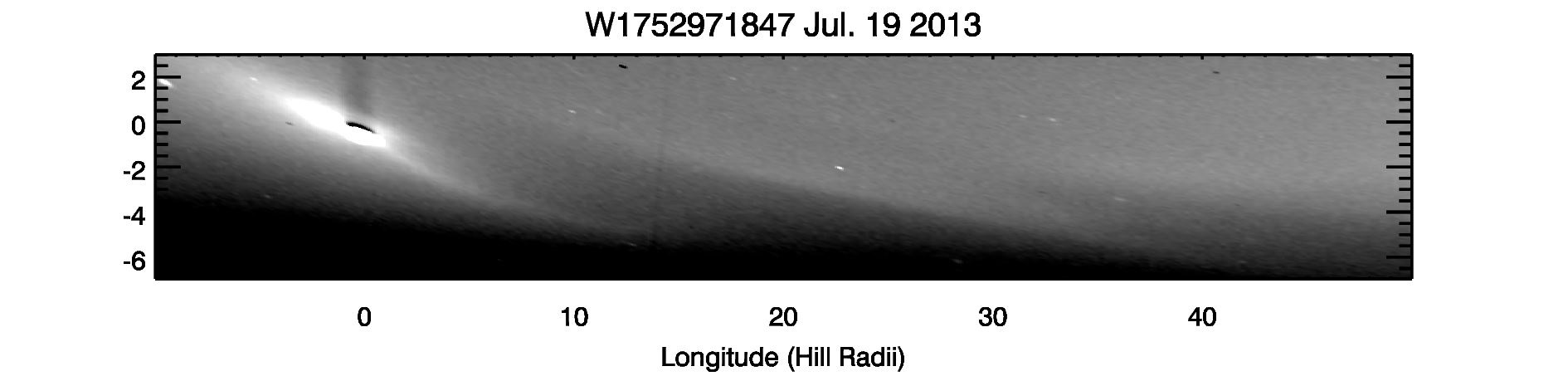}}
\caption{Re-projected images of the E-ring structures around Enceladus. Note that both longitude and radius are measured in Hill Radii and the origin is located at Enceladus. Unlike Figure~\ref{m15}, these images are neither spatially filtered nor brightness inverted (although they are independently stretched), and so regions of higher particle density appear brighter than the background. For this analysis we are particularly interested in the feature centered at a radius of  -2 Hill Radii and a longitude of +20 Hill Radii.}
\label{rpj}
\end{figure*}

In order to compare the locations and brightnesses of the bright streaks seen in Figure~\ref{m15} with the expected properties of rebound wakes, we re-processed the data from both these images. Table~\ref{images used} provides a brief summary of the observation parameters for both of these images, including the phase angle, resolution and the inertial longitudes (measured relative to the ascending node of the rings on the J2000 equator plane) of the Sun, Enceladus, and Enceladus' orbital pericenter. We did examine several other images of this region, including those analyzed by \citet{Mitchell15}, but found that those images either did not capture a wide enough range of longitudes or had much lower signal-to-noise, and so they were not included in this investigation.

The raw data for both images were calibrated using the standard CISSCAL routines \citep{Porco04, West10}, and were geometrically navigated based on the locations of known stars in the field of view. The data from both images were then re-projected to produce maps of the ring's projected brightness as a function of radius and inertial longitude. Finally, we converted the relevant coordinates to radial and azimuthal distances from Enceladus in units of Hill Radii  in order to facilitate comparisons with the predicted wake locations, 

Figure~\ref{rpj} shows the resulting maps of the E ring in the vicinity of Enceladus on a common scale. Note that these images are neither brightness-inverted nor high-pass filtered, so regions of higher density appear as bright streaks in these images. The feature \citet{Mitchell15} identified as the interior tendril appears as a diagonal streak just below Enceledus, while the first leading tendril is the feature centered around 2 Hill Radii interior to Enceladus' orbit and 20 Hill Radii ahead of the moon. Consistent with prior findings by \citet{Mitchell15}, the observed locations of these bright bands are inconsistent with predicted locations of passing satellite wakes shown in Figure~\ref{models}. However, the first and most prominent rebound wake extends between 10 and 20 Hill Radii in azimuth and between -1 and -3 Hill Radii in Radius, and so this feature occurs quite close to the  observed location of a bright band in both images. 

In order to more precisely  compare the observed locations of these bright bands with the expected locations of the first rebound wake, we used the following procedures to trace the locations of the bright streaks in both maps. First, we generated a high-pass filtered version of each map by subtracting a version of the relevant map smoothed over a spatial scale of 4000 km, {comparable to that used in Figure~\ref{hillsim4}.\footnote{This smoothing scale is larger than the one used for the simulations in Figure~\ref{hillsim2} because the background brightness variations in the ring are on a larger scale due to the broader range of eccentricities.}  }As shown in the top panels of Figures~\ref{hillcomp2} and~\ref{hillcomp}, this filtering removed large-scale brightness trends but preserved the signals from the bands. Next, for each longitude in each map, we identified the radius of the brightness maximum within a selected range of radii that contained the interior and leading tendrils. The middle panels of Figures~\ref{hillcomp2} and~\ref{hillcomp} show the resulting wake locations overlaid on the high-pass-filtered images, demonstrating that this process properly traces the location of the structures. We also quantified the brightness variations  associated with these structures in terms of  the peak fractional brightness variation at each longitude for each feature in both images. The peak fractional brightness variation is simply the ratio of the brightness of the peak in the high-pass-filtered image to the  brightness of the unfiltered image, and these numbers are plotted in Figure~\ref{bright}.

\section{Comparisons between observed bright streaks and predicted properties of rebound Wakes}
\label{results}

\subsection{Locations of bright streaks overlap with the expected positions of rebound wakes}

\begin{figure}
    \centering
    \resizebox{3.5in}{!}{\includegraphics{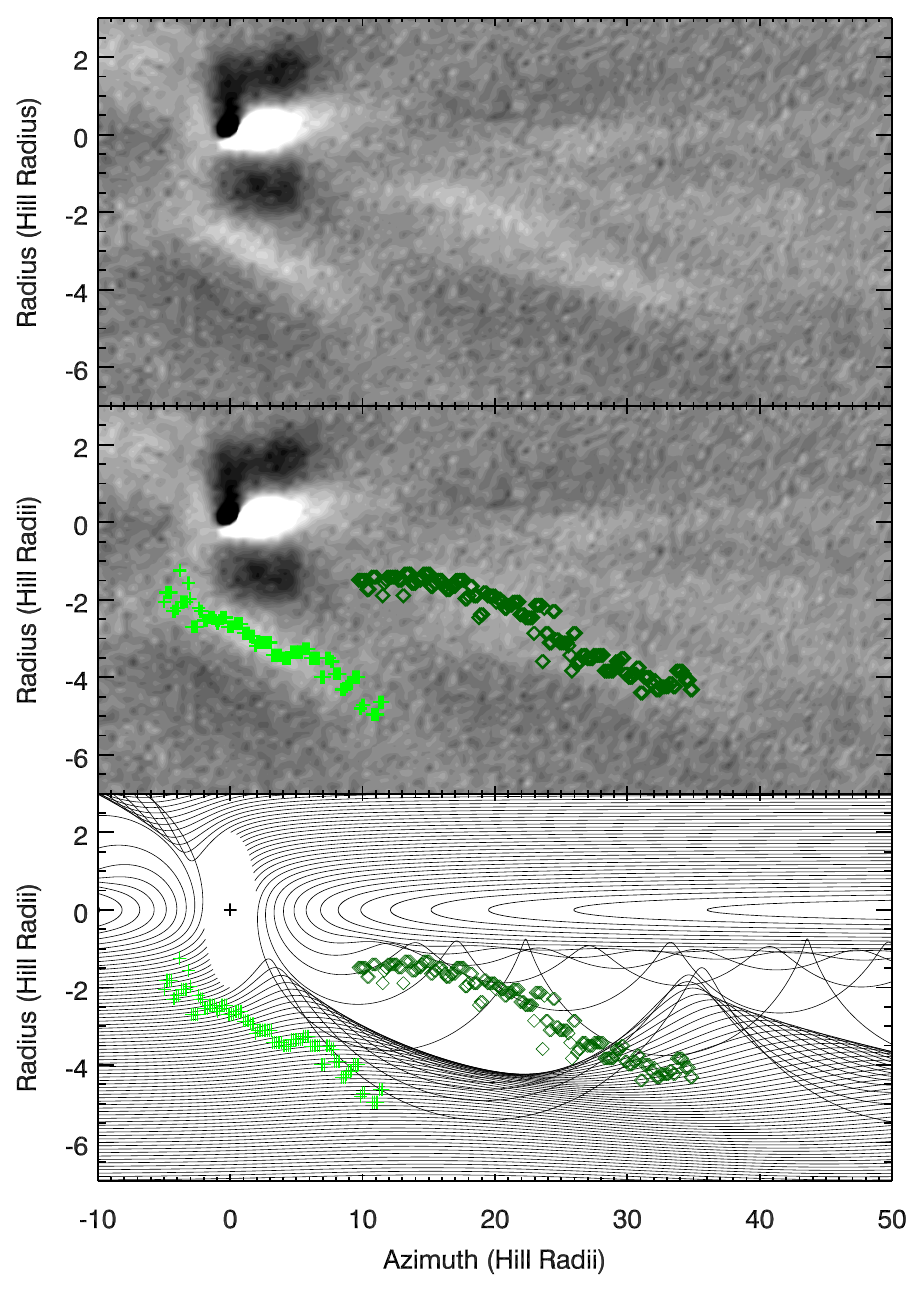}}
    \caption{Comparisons of the observed positions of features in image W1537013154 with those expected from passing and rebound wakes. The top panel shows a high-pass filtered version of the reprojected image, where the brightness variations can be more clearly seen (brightness scale runs from $\pm$10\%). The second panel shows the same image with the estimated locations of the brightness maxima marked as lime crosses and green diamonds. The bottom panel overlays these same positions with the trajectories predicted by Hill's Equations. Note that the structure connected to Enceladus is displaced several Hill Radii to the left of the first passing wake, the second brightness maxima lie very close to the expected location of the first rebound wake, although it does appear to extend beyond that wake.}
    \label{hillcomp2}
\end{figure}

\begin{figure}
    \centering
    \resizebox{3.5in}{!}{\includegraphics{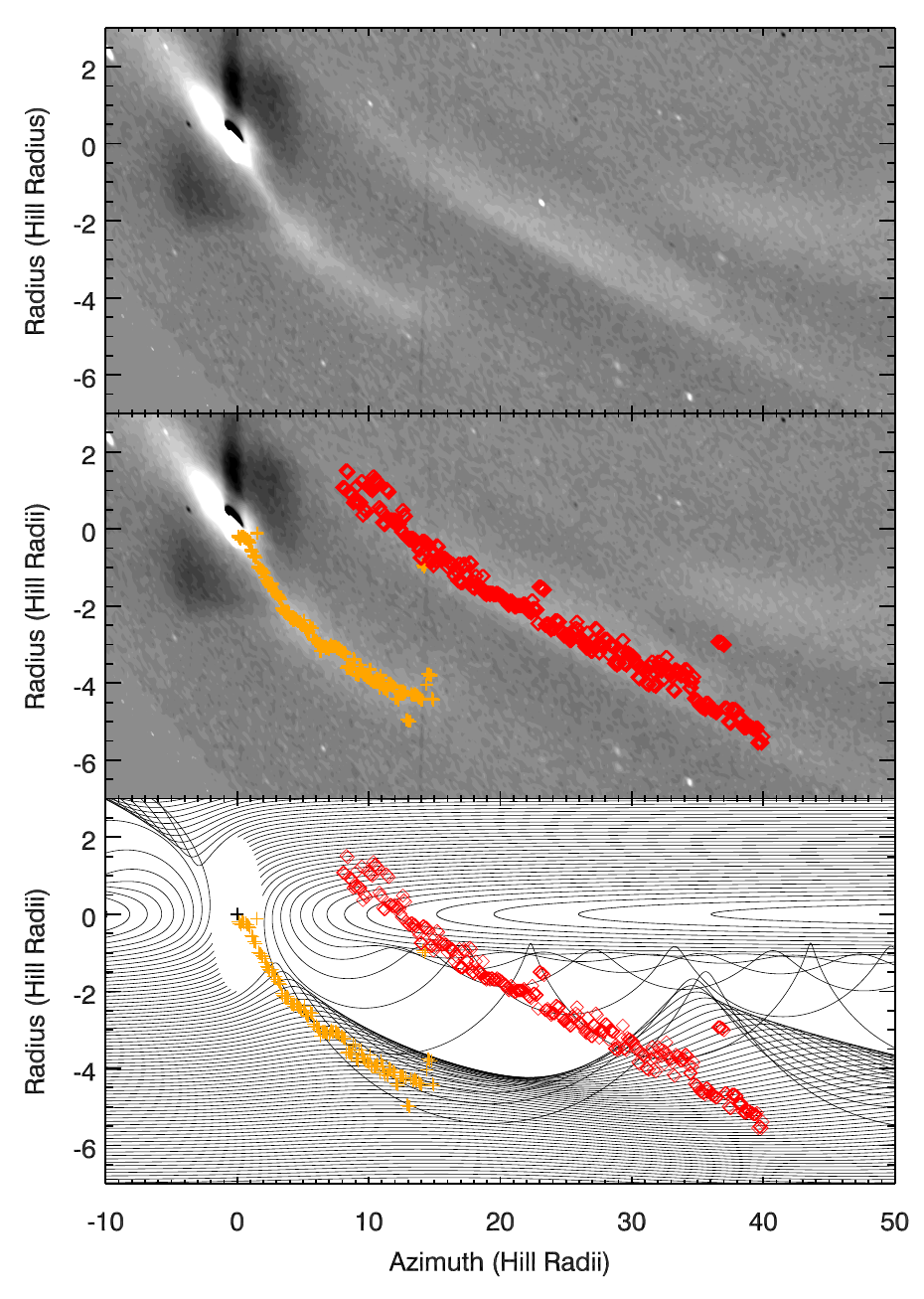}}
    \caption{Comparisons of the observed locations of features in image W1752971847 with those expected from passing and rebound wakes. The top panel shows a high-pass filtered version of the reprojected image, where the brightness variations can be more clearly seen (brightness scale runs from $\pm$10\%). The second panel shows the same image with the estimated locations of the brightness maxima marked as orange crosses and red diamonds. The bottom panel overlays these same positions with the trajectories predicted by Hill's Equations. Note that the structure connected to Enceladus is displaced a couple of Hill Radii to the left of the first passing wake, the second brightness maxima lie very close to the expected location of the first rebound wake, although it does appear to extend beyond that wake.}
    \label{hillcomp}
\end{figure}

The bottom panels of Figures~\ref{hillcomp2} and~\ref{hillcomp} compare the observed positions of the bright streaks in the two maps with the predicted locations for both the passing and rebound wakes.

If we first consider the bright streaks directly interior to Enceladus, it is clear that their locations are inconsistent with those expected for any satellite wakes. For image W1537013154, the bright band lies well behind the expected location of the wake, and even extends slightly behind Enceladus itself. For W1725971847 the bright feature is closer to the expected location of the wake,  but is still roughly 2 Hill Radii behind the wake's location. Furthermore, the fact that this feature's position shifts between the two observations is inconsistent with the expected behavior of satellite wakes. Hence, these structures are most likely tendrils consisting of material recently launched from Enceladus, whose precise positions can change depending on the locations of the active sources, the assumed launch velocity of the particles, and the observation geometry \citep{Mitchell15}.

At the same time, it is important to note that there are no obvious brightness enhancements at the predicted locations of the first passing wake in image W1537013154 (see Figure~\ref{hillcomp2}), even though the observed tendril is well displaced from the expected wake position. The lack of any observable wake signatures implies that there are not sufficient particles on suitable orbits to form detectable interior passing wakes in the E ring \cite[cf.][]{Mitchell15}.

However, if we instead consider the bright streaks lying between 10 and 30 Hill Radii ahead of the moon, the situation is more complicated. The material more than 20 Hill Radii in front of Enceladus falls at the predicted locations of the leading tendrils \citep{Mitchell15}, and these parts of the streaks extend beyond the expected locations of rebound wakes, so the distal parts of these streaks are almost certainly tendrils. However, the portions of these bright bands that lie around 15-20  Hill Radii in front of Enceladus and around 2 Hill Radii interior to Enceladus' orbit are located almost exactly where the first rebound wake should occur.  Furthermore, in image W1537103154, there appears to be two overlapping streaks in this region (see Figure~\ref{hillcomp2}), one situated between 10 and 20 Hill Radii in azimuth and around -1.5 Hill Radii in radius, while the other extends between 20 and 30 Hill Radii in azimuth and between -3 and -4 Hill Radii in radius. The lower feature extends over a range of radii similar to the tendril below the moon, and corresponds to the portion of this feature that was reproduced reasonably well as a tendril by \citet{Mitchell15} (see Figure~\ref{m15}). By contrast, the upper part of this band was the portion of this feature that those authors were not able to reproduce in their simulations. Thus it is reasonable to consider the possibility that the lower feature is the actual leading tendril, while the upper feature is a rebound wake. The structure in image W1752971847 cannot be so easily separated into two distinct components, but the fact that the portion of the band between 10 and 20 Hill Radii is at very similar locations in both images is  consistent with the expected behavior of a satellite wake.

One further complication in interpreting this feature is that the part of the feature closest to Enceladus is not in exactly the same location in the two images. In the earlier image W1537013154, the end of the band curves inwards towards the planet and so falls interior to the expected wake location between 10 and 15 Hill Radii in front of the moon. By contrast in the W1752971847 image the bright band curves outwards, and so falls exterior to the expected wake location in this region and in fact extends all the way to Enceladus' orbit.  These variations pose challenges to both the tendril and rebound wake models, and will likely require more detailed simulations to explain.

\begin{figure*}
    \centering\resizebox{6.5in}{!}{
    \includegraphics{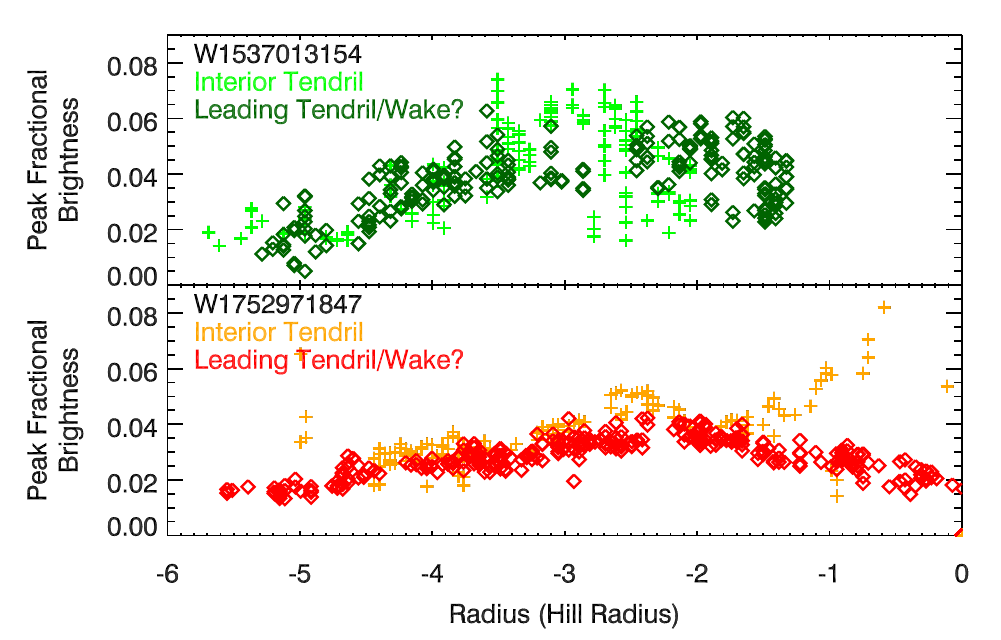}}
    \caption{Plots showing the peak fractional brightness of the bright streaks in the two images as a function of radial displacement from Enceladus' orbit. Note the color code matches that used in Figures~\ref{hillcomp2} and~\ref{hillcomp}.}
    \label{bright}
\end{figure*}

\subsection{The brightness of these features could be consistent with the E-ring's orbital element distribution}

The other main argument made by \citet{Mitchell15} against the idea that these features could be satellite wakes is that such wakes only form among particles that have nearly the same eccentricity and pericenter as Enceladus, which is a small fraction of the particles in the E ring. As discussed in Section~\ref{theory} above, particles with epicylic motions relative to Enceladus larger than  around 1 Hill Radius are unlikely to form wakes because their radial motions will prevent them from producing the observed small-scale density variations. We therefore need to evaluate whether the brightness of these features is inconsistent with reasonable estimates of the orbital element distribution of the E ring.

As shown in Figure~\ref{bright}, these features are actually quite subtle, representing variations in the background E-ring density of only about 5\%.  The fractional brightness variations of the relevant feature in the images are therefore roughly comparable to the fractional density variations shown in the simulations in Figures~\ref{hillsim2} and~\ref{hillsim4}. Thus the features in the wakes could potentially be wake signatures so long as a large enough fraction of the particles in this part of the E ring have the requisite orbital elements. 

{As mentioned in Section~\ref{theory} above, the best constrained aspect of the particles' orbital properties is the eccentricity distribution, and the simulations discussed above demonstrate that this broad eccentricity distribution allows for wakes with the required intensity to form. Also, the distributions of pericenter locations and inclinations are also unlikely to strongly affect the intesnity of the wake signatures.} While there are longitudinal variations in the shape of the E-ring profile that could be due to asymmetries in the pericenter direction, these are relatively subtle and are tied to the direction of the Sun rather than the location of Enceladus \citep{Hedman12, Ye16}. Hence there are probably not strong asymmetries in the distribution of  pericenter locations. In addition, even though  the E-ring particles do have a range of orbital inclinations, this parameter is strongly correlated with the eccentricity and so it is reasonable to expect that particles with low eccentricities also have low inclinations. Hence it is probably reasonable to neglect orbital inclinations for this preliminary investigation.

{By contrast, the particles' orbital semi-major axis distribution certainly strongly affects the intensity of these sorts of wake signatures.} Only particles orbiting between 1 and 1.5 Hill Radii exterior to Enceladus can produce the required rebound wake, so we need a substantial fraction of the particles on low-eccentricity orbits to have their semi-major axes in this range. At the moment, there are no measurements of the semi-major axis distribution of particles close to Enceladus' orbit with high enough fidelity to reliably estimate the fraction of particles that would be on these wake-producing orbits. 

{In fact, the relative intensities of small-scale E-ring structures like these wakes probably provide the best opportunities to constrain the detailed shape of the E-ring particles' orbital parameter distribution.  For example, the lack of obvious signatures of passing wakes indicates that there are comparatively few particles on low-eccentricity orbits with semi-major axes between 2.5 and 4 Hill radii from Enceladus' orbit, which could support the idea that a substantial fraction of the E-ring particles on low-eccentricity orbits are on horseshoe orbits.  On the other hand, the same  simulations in Figures~\ref{hillsim2} and~\ref{hillsim4}  that have a strong rebound wake also tend to have a $\subset$-shaped feature that traces the average trajectory of the same particles responsible for the wake. This $\subset$-shaped feature is not clearly visible in the images, which could place an upper limit on the fraction of particles on orbits that could produce a rebound wake. However, initial investigations show that the relative strength of the wake signature and the $\subset$-shaped feature depends on the details of both the semi-major axis and eccentricity distributions. A broad range of orbital parameter distributions will therefore probably need to be investigated to ascertain whether particle populations capable of produced detectable rebound wakes are consistent with other aspects of the E-ring's structure.}

\section{Discussion and Conclusions}
\label{discussion}

The above analysis indicates that while there is no evidence for passing satellite wakes in the E ring, rebound wakes could contribute to the bright features observed 10-20 Hill Radii in front of Enceladus. Even though the distal ends of these features are most likely tendrils of material recently launched from Enceladus \citep{Mitchell15}, the bright bands around 15 Hill Radii in front of Enceladus are in the right locations to be due to pre-existing E-ring particles following horseshoe orbits. Furthermore, these features are sufficiently subtle that they could be generated by the small fraction of E-ring particles that have suitable near-circular orbits.  Properly quantifying and/or limiting the contributions from rebound wakes to these features will require additional numerical simulations of plume and E-ring material, as well as more detailed analysis of the available data.

{While such detailed investigations are beyond the scope of this initial study,  we can at least point out a potential mechanism that would allow a large fraction of the particles on low-eccentricity orbits to approach Enceladus on trajectories that could produce the rebound wake signature.} The tendrils found in the \citet{Mitchell15} simulations correspond to material with semi-major axes between those of horseshoe and passing orbits, or around 2 Hill Radii interior and exterior to Enceladus' orbit. After being launched from Enceladus, these particles are expected to migrate outward due to plasma drag and other non-gravitational processes \citep{Burns76, Grun84, Horanyi92, Juhasz02, Juhasz04, Juhasz07, Horanyi08}. This outward migration will cause material initially launched Saturn-wards from Enceladus to approach and then move past Enceladus'  orbit, so while these particles would first drift ahead of the moon, they should eventually reverse direction and drift back towards the moon \citep{Murray94, Hedman13}. By symmetry, these particles would re-encounter Enceladus when their semi-major axes is about 2 Hill Radii exterior to Enceladus' orbit. This would naturally cause many of these particles to approach Enceladus along semi-major axes close to those needed to generate rebound wakes. Of course, this process could also deliver particles onto orbits that could produce passing wakes exterior to Enceladus' orbit. Quantitatively evaluating the expected strength of these different classes of wakes willrequire sophisticated numerical simulations including a variety of non-gravitational forces.

A larger suite of numerical simulations of material launched from Enceladus would also clarify whether this material can ever produce features as close to Enceladus as observed in these images. The fact that the \citet{Mitchell15} simulations did not produce tendrils extending as close to Enceladus as the observations suggests that, if these features are not rebound wakes, then their morphology will place strong constraints on the particle launch velocity distribution from Enceladus.  {At the same time, simulations of various populations of E-ring material are needed to determine which orbital element distributions could produce structures comparable in brightness to those observed,  while at the same time avoiding producing other structures in the rings (like signals from the passing wakes or $\subset$-shaped patterns) that are too intense to be consistent with the data. Since only particles with a narrow range of orbital parameters can produce each of these structures, determining the fraction of particles with these properties should place strong constraints on the evolution rates of the particles' semi-major axes and eccentricities.}

\pagebreak

\section*{Acknowledgements}

This work was supported in part by NASA Cassini Data Analysis Program grant number NNX15AQ67G. The authors wish to thank C. Mitchell and C. Porco for their helpful discussions.

%\bibliographystyle{icarus}
%\bibliography{ering.bib} 

\end{document}